\documentclass{pinchcr}

\usepackage{amsmath}
\usepackage{amssymb}
\usepackage[english]{babel}
\usepackage[superscript]{cite}
\usepackage{calrsfs}
\usepackage{graphics}
\usepackage{graphicx}
\usepackage{setspace}

\DeclareMathOperator{\sech}{sech}

\title{White-paper Draft}
\author{J. Schwinger}
\affiliation{Affiliation1}
\author{A. Author2 and A. Author3}
\affiliation{Affiliation for Author2 and Author3}
\authors{2}

\makeatletter

\newcommand{\Rmnum}[1]{\expandafter\@slowromancap\romannumeral #1@}

\makeatother

\begin{document}

%\begin{center}
%	\noindent\textbf{\LARGE{APS-MARCH MEETING - BOSTON - 2009}} \\
%	\textbf{\LARGE{Abstract: T70.00285}}
%\end{center} 

\begin{center}
	\textbf{\newline STATISTICAL THERMODYNAMICS OF THE ``DICE LATTICE"}
\end{center}
	
	\par\makebox[4cm]{}\Large{Norman J.M. Horing} \par
	\makebox[4cm]{}\Large{Department of Physics} \par
	\makebox[4cm]{}\Large{Stevens Institute of Technology} \par
	\makebox[4cm]{}\Large{Hoboken, NJ 07030, USA} \par	
	\makebox[4cm]{} \par
	\makebox[4cm]{}\Large{M.L. Glasser} \par
	\makebox[4cm]{}\Large{Department of Physics} \par
	\makebox[4cm]{}\Large{Clarkson University} \par
	\makebox[4cm]{}\Large{Potsdam, NY 13699, USA} \par
	\makebox[5.2cm]{}\Large{and} \par
	\makebox[4cm]{}\Large{Jay D. Mancini} \par
	\makebox[4cm]{}\Large{Department of Physical Science} \par
	\makebox[2.5cm]{}\Large{Kingsborough Community College, CUNY} \par
	\makebox[4cm]{}\Large{Brooklyn, New York 11235} \par
	\makebox[4cm]{} \par
	\makebox[4cm]{}\Large{August 9, 2020} \par

	\makebox[4cm]{} \par
	
	%\makebox[4cm]{}\Large{May 9, 2020} \newline 
	
	\begin{center}
		\textbf{ABSTRACT}
	\end{center}
	
%\vspace*{\fill}
In this work we analyze the statistical thermodynamics of "Dice" lattice carriers employing a Green's function formulation to examine the grand potential, Helmholtz free energy, the grand and ordinary partition functions and entropy. This facilitates the calculation of the specific heat, and all evaluations are carried out for both the degenerate and nondegenerate statistical regimes. \newline

\newpage
\noindent \textbf{1. INTRODUCTION}
	
This work addresses the fundamental statistical thermodynamic properties of "Dice" lattice$^{1,2}$ carriers. This system is a two-dimensional psuedospin 1 lattice, and a recent study of it by Malcolm and Nicol$^1$ analyzed its dynamic, nonlocal polarizability underlying its plasmon spectrum and static shielding features. The "Dice" lattice is a recent addition to the list of Dirac materials, which have low energy spectra and associated Hamiltonians linearly proportional to momentum. Other such Dirac materials include Group VI Dichalcogenides$^3$, Topological Insulators$^4$, Silicene$^5$, and, of course, Graphene$^{6-11}$: It was the discovery of the exceptional electrical conduction and sensing properties of Graphene about 15 years ago that focused attention on Dirac materials and brought Geim and Novoselov the 2010 Nobel Prize: Dirac materials are currently under investigation in laboratories worldwide for their potential to succeed Silicon as the basis for the next generation of computers and electronics. Intellectual interest in them is further heightened by the fact that their Hamiltonians and spectra are similar to that of relativistic electrons/positrons.

In Section 2, the Grand Potential, Helmholtz Free Energy and the grand and ordinary partition functions are formulated in terms of the retarded Green's function of Dice lattice carriers, and the Green's function is determined explicitly in frequency/momentum representation. Section 3 presents the determination of the Grand Potential as a function of temperature in both the degenerate regime (with the approach to the zero temperature limit) and the nondegenerate regime. In Section 4 we exhibit the results for entropy and specific heat at constant volume in both the degenerate and nondegenerate regimes. The specific heat is of particular interest as a measure of the ability of the material to assist in the management of dissipated heat, an issue of importance in electronic device operation and transport. Specific heat also plays an important role in a standard characterization technique employed to understand the underlying physics of the materials, as has been emphasized by Stewart$^{12}$ and Geballe's group$^{13-17}$.
\newpage
\noindent \textbf{2. GRAND POTENTIAL AND THE GREEN'S \newline FUNCTION}
\newline
\indent Our formulation of the problem is focused on the Grand Potential, $\Omega$, of the ``Dice" lattice,
%\Large
\begin{equation}
	\Omega \equiv F-\mu N = -\kappa_B T' \ln Z = -\kappa_B T' \sum_{E_\gamma} \ln (1+e^{-\beta[E_\gamma - \mu]}) \text{,} \tag{2.1}
\end{equation}
where $F$ is the Helmholtz Free Energy, $N$ is particle number, $\kappa_B$ is the Boltzmann constant, $T'$ is Kelvin temperature, $\mu$ is the chemical potential, $Z$ is the grand partition function, $\beta = 1/\kappa_B T'$ and $E_\gamma$ represents the particle energy spectrum (which is summed over). A.H. Wilson$^{18}$ reformulated the expression for $\Omega$ in terms of the $E_\gamma$-summand function, $B(E_\gamma)$, on the right of Eq.(2.1) as an inverse Laplace transform (c represents the inverse Laplace transform integration contour):
\begin{equation}
	B(E_\gamma) = -\kappa_B T' \ln(1+e^{-\beta[E_\gamma - \mu]}) = \int_{c} \frac{ds}{2 \pi i} e^{sE_\gamma} p(s) \text{,} \tag{2.2}
\end{equation}
with $p(s)$ as the Laplace transform of $B(E)$,
\begin{equation}
	p(s) = \int_{0}^{\infty} dE e^{-sE} B(E) \: \text{.} \tag{2.3}
\end{equation}
Wilson employed a special case of the convolution theorem for Laplace transforms$^{19}$ to show that
\begin{equation}
	\Omega = F -\mu N = \int_{0}^{\infty} dE \int_{c} \frac{ds}{2 \pi i} e^{Es} \frac{\widehat{Z} (s)}{s^2} \int_{c} \frac{ds'}{2 \pi i} e^{Es'} {s'}^2 p(s') \text{,} \tag{2.4}
\end{equation}
where $\widehat{Z}(s)$ is the ordinary (not ``grand") partition function. Moreover,
\begin{equation}
	\int_{c} \frac{ds'}{2 \pi i} e^{Es'} {s'}^{2} p(s') = \frac{\partial^2}{\partial E^2} \int_{c} \frac{ds'}{2 \pi i} e^{Es'} p(s') = \frac{\partial^2 B(E)}{\partial E^2} = \frac{\partial f_0 (E)}{\partial E} \text{,} \tag{2.5}
\end{equation}
where $f_0 (E)$ is the Fermi-Dirac distribution function, so the calculation of $\Omega$ is conveniently reformulated in terms of the Fermi distribution at arbitrary temperature $T'$ and the ordinary partition function:
\begin{equation}
	\Omega = F - \mu N = \int_{c} \frac{ds}{2 \pi i} \frac{\widehat{Z}(s)}{s^2} \int_0^\infty dEe^{Es} \frac{\partial f_0 (E)}{\partial E} \text{.} \tag{2.6}
\end{equation}
A particular advantage of dealing with the ordinary partition function $\widehat{Z} (s)$ is that it may be conveniently obtained directly from the trace of the retarded Green's function $G_{T>0}^{ret} (\vec{x}, \vec{x'}; T)$ at positive time difference $T>0$ as$^{20} (\beta = 1 / \kappa_B T'$, where $T'$ represents Kelvin temperature$)$ 
\begin{equation}
	\widehat{Z} (\beta) = \text{trace}(e^{-\beta H}) = \int d^2 x \:\: Tr(\hat{i} G_{T>0}^{\text{ret}} (\vec{x},\vec{x}; T \rightarrow -i\beta)) \text{,} \tag{2.7}
\end{equation}
where the 2D $d^2x$-integral provides an area factor for a uniform sheet. Here, we have used the fact that $e^{-iHT}$ embedded in the structure of the Green's function is the time translation operator ($Tr$ denotes the pseudospin trace of the matrix Green's function). $H$ is the Hamiltonian and for the Dice lattice in momentum ($\vec{K}$) representation it is given by the $3\times 3$ pseudospin-1 matrix$^{1,2}$ 
\begin{equation}
	H = \alpha\left[\begin{array}{ccc}
	0&K_-&0 \\
	K_+&0&K_- \\
	0&K_+&0\end{array}\right] \:\: \text{where } K_\pm = K_x \pm iK_y \text{,} \tag{2.8}
\end{equation}
and $\alpha = \hbar v /\sqrt{2}$ with $v$ as the Fermi velocity. The determination of the Dice lattice Green's function $G^\text{ret}$ is done in frequency-momentum representation using the matrix equation ($I$ is the $3\times 3$ unit matrix)
\begin{equation}
	(I \omega - H)G^{\text{ret}} = I \text{,} \tag{2.9}
\end{equation}
with the resulting matrix elements of $G^{\text{ret}}$ given by:
\begin{equation}
	G^\text{ret} (\vec{K},\omega) = \left[\begin{array}{ccc}
	G_{11}&G_{12}&G_{13} \\
	G_{12}^*&G_{22}&G_{23} \\
	G_{13}^*&G_{23}^*&G_{33}\end{array}\right] \text{,} \tag{2.10}
\end{equation}
where we note that $G_{ij} = G_{ji}^*$ due to hermiticity of the Green's function (and Hamiltonian). Here,
\begin{align}
	G_{11} &= \frac{1}{\omega} \:  \frac{\omega^2-\alpha^2K^2}{\omega^2-2\alpha^2K^2} \tag{2.11} \\
	G_{22} &= \frac{\omega}{\omega^2-2\alpha^2K^2} \tag{2.12} \\
	G_{33} &= \frac{1}{\omega} + \frac{\alpha^2 K^2 / \omega}{\omega^2-2\alpha^2K^2} = G_{11} \tag{2.13} \\
	G_{12} &= \frac{\alpha K_-}{\omega^2-2\alpha^2K^2} \: ; \: G_{12}^* = \frac{\alpha K_+}{\omega^2-2\alpha^2K^2}, \tag{2.14} \\
	G_{23} &= \frac{\alpha K_-}{\omega^2-2\alpha^2K^2} \: ; \: G_{23}^* = \frac{\alpha K_+}{\omega^2-2\alpha^2K^2}, \tag{2.15} \\
	G_{13} &= \frac{\alpha^2 K_-^2 / \omega}{\omega^2-2\alpha^2K^2} \: ; \: G_{13}^* = \frac{\alpha^2 K_+^2 / \omega}{\omega^2-2\alpha^2K^2}, \tag{2.16}
\end{align}
where $\omega \rightarrow \omega + i0^+$ for the retarded Green's function. To obtain the ordinary partition function, $\widehat{Z}$, it is necessary to obtain the trace of the functional form of the Green's function in direct time representation with the substitution of its positive time difference argument $T$ replaced by $T \rightarrow -i\beta^{20}$. This also involves a change from momentum representation to position representation for the relative positional argument $\vec{R} = \vec{x} - \vec{x} = 0$. Thus, we have
\begin{equation}
	\widehat{Z} (\beta) = (\text{area})i \int \frac{d^2 K}{(2 \pi)^2} \left[\int \frac{d \omega}{2 \pi}e^{-i \omega T}Tr G^{\text{ret}}(\vec{K},\omega)\right]_{T \rightarrow -i\beta}. \tag{2.17}
\end{equation}
\newpage
\noindent \textbf{3. EVALUATION OF THE GRAND POTENTIAL}

Employing Eqns. (2.11-2.13) to determine the Dice lattice Green's function trace, the result may be exhibited in terms of its frequency/energy poles as
\begin{equation}
	TrG^{\text{ret}}(\vec{K},\omega) = \frac{1}{\omega} + \sum_\pm \frac{1}{\omega \pm \sqrt{2 \alpha^2 K^2}} \text{,} \tag{3.1}
\end{equation}
and then ($T>0$ and $\omega \rightarrow \omega + i0^+$)
\begin{align}
	&\int_{-\infty}^{\infty} \frac{d\omega}{2\pi} e^{-i\omega T} TrG^{\text{ret}}(\vec{K},\omega) \nonumber \\ 
	&= -i\left(1+\sum_{\pm} \exp( \pm i\sqrt{2\alpha^2 K^2}T)\right) \text{.} \tag{3.2}
\end{align}
With the substitution $T \rightarrow -i\beta$, we obtain the ordinary partition function $\widehat{Z}(\beta)$ (per unit area) as
\begin{equation} 
\widehat{Z}(\beta) = \int \frac{d^2 K}{(2\pi)^2} \left(1+\sum_{\pm} \exp( \pm \sqrt{2\alpha^2 K^2} \beta)\right) \text{.} \tag{3.3}
\end{equation}
Bearing in mind that the underlying band structure departs from its low energy linear approximation at some maximum crystal momentum value $K_m$, we introduce that as an upper cutoff on the $K$-integral, whence
\begin{equation}
	\widehat{Z} (\beta) = \frac{1}{2\pi} \int_{0}^{K_m} dK \cdot K\left(1+\sum_{\pm} \exp( \pm \sqrt{2} \alpha \beta K)\right) \text{,} \tag{3.4}
\end{equation}
which yields the result
\begin{equation}
	\widehat{Z} (\beta) = \frac{K_m^2}{4 \pi} - \frac{\cosh (\sqrt{2} \alpha \beta K_m)}{2\pi \alpha^2 \beta^2} + \frac{K_m \sinh (\sqrt{2} \alpha \beta K_m)}{\sqrt{2} \pi \alpha \beta} + \frac{1}{2\pi \alpha^2 \beta^2} \text{.} \tag{3.5}
\end{equation}
To examine the degenerate regime we employ Eq. (2.6) jointly with
\begin{equation}
	\frac{\partial f_0 (E)}{\partial E} = \frac{-\beta}{4} \sech^2 \left(\frac{[E-\mu] \beta}{2}\right) \tag{3.6}
\end{equation}
and introduce the variable $z = [E-\mu]\beta / 2$, so that
\begin{equation}
	\int_{\partial}^{\infty} dE \: e^{Es} \frac{\partial f_0 (E)}{\partial E} = -\frac{1}{2} e^{s\mu} \int_{-\mu \beta / 2}^{\infty} dz \: e^{2sz/\beta} \sech^2 (z) \text{.} \tag{3.7}
\end{equation}
In the degenerate regime $\mu \beta \rightarrow \infty$, so the lower limit of the $z$-integral may be taken as $-\mu \beta / 2 \rightarrow - \infty$, with the result$^{21}$
\begin{equation}
	\int_{0}^{\infty} dE \: e^{Es} \frac{\partial f_0 (E)}{\partial E} = -\frac{\pi}{\beta} \: \frac{se^{s\mu}}{\sin (\frac{\pi s}{\beta})} \text{.} \tag{3.8}
\end{equation} 
Correspondingly, Eq. (2.6) yields the grand potential per unit area as $(s^{'} \equiv \frac{s}{\beta})$
\begin{equation}
	\Omega = F - \mu n = \frac{-\pi}{\beta} \int_z \frac{ds^{'}}{2\pi i} \frac{e^{s^{'}\beta\mu} \widehat{Z}(\beta s^{'})}{s^{'} \sin (\pi s^{'})} \text{,} \tag{3.9}
\end{equation}
and it is convenient to employ $\widehat{Z}(\beta s^{'})$ from Eq. (3.4), writing $(v \equiv \sqrt{2} \alpha)$
\begin{equation}
	\widehat{Z}(\beta s^{'}) = \frac{K_m^2}{4\pi} + \sum_\pm \int_0^{K_m} \frac{dK}{2\pi} Ke^{\pm v\beta K s^{'}} \text{.} \tag{3.10}
\end{equation}
In the ensuing $s^{'}$-integral of Eq. (3.9) to obtain $\Omega$, we exponentiate the integrand denominator factor $1/s^{'} = \beta \int_0^\infty dx e^{-s'\beta x}$ so that a particular term with energy $E_\gamma$ contributes as 
\begin{equation}
	\int_c \frac{ds^{'}}{2\pi i} \frac{e^{s'\beta E_\gamma}}{s' \sin(\pi s')} = \beta \int_0^\infty dx \int_c \frac{dz}{2\pi i} \frac{e^{z\beta [E_\gamma / \pi - x]}}{\sin(z)} \text{.} \tag{3.11}
\end{equation} 
Noting that the contour of $z$-integration along c is a straight line from $z = -i\infty + 0^+$ to $+i\infty + 0^+$, we consider closing the contour with a parallel line $(c')$ from $i\infty - \pi^+$ to $-i\infty - \pi^+$ on which $dz_{c'} = -dz_c$ and $\sin(z_{c'}) = -\sin(z_c)$.$^{22}$ Moreover, the closed contour integrand $\oint = \int_c + \int_{c'}$ has the residue ``1" at $z = 0$, so that 
\begin{equation}
	\begin{split} 
	\oint \frac{dz}{2\pi i} \cdots = \int_c \frac{dz_c}{2\pi i} \cdots + \int_{c'} \frac{dz_{c'}}{2\pi i} \cdots = \left( 1 + e^{-\pi\beta [E_\gamma / \pi - x]} \right) \\ \times \int_c \frac{dz}{2\pi i} \frac{e^{z\beta [E_\gamma / \pi - x]}}{\sin(z)} = 1 \text{.}
	\end{split} \tag{3.12}
\end{equation}
Consequently, the $x$-integration of Eq. (3.11) is given by 
\begin{equation}
	\int_0^\infty dx \frac{1}{1 + e^{-\pi \beta [E_\gamma / \pi - x]}} = \frac{1}{\pi \beta} \ln (1 + e^{\beta E_\gamma}) \text{,} \tag{3.13}
\end{equation}
and for the degenerate regime under consideration, we obtain the Grand Potential $\Omega$ as
\begin{equation}
	\Omega = \frac{-K_m^2}{4\pi\beta} \ln (1 + e^{\beta\mu}) - \frac{1}{2\pi\beta} \sum_\pm I_\pm \text{,} \tag{3.14} 
\end{equation}
where
\begin{equation}
	I_\pm = \int_0^{K_m} dK K \ln (1+ e^{\beta [\mu \pm vK]}) \text{.} \tag{3.15}
\end{equation}
Since $K_m >> \mu/v >> k_B T/v$ in the degenerate regime, $I_+$ is readily approximated as
\begin{equation}
	I_+ \equiv \frac{\beta\mu K_m^2}{2} + \frac{\beta v K_m^3}{3} \text{.} \tag{3.16}
\end{equation}
$I_-$ requires more careful analysis in the low wave number region of the $K$-integration: Integrating by parts,
\begin{equation}
	I_- = \frac{K_m^2}{2} \ln (1 + e^{\beta (\mu - vK_m)}) + \frac{\beta v}{2} \int_0^{K_m} dK K^2 f_0 (vK) \text{,} \tag{3.17}
\end{equation}
and integrating by parts again we obtain the Grand Potential in the degenerate regime as$^{23}$
\begin{equation}
	\begin{split}
	\Omega = \frac{-K_m^2}{4\pi \beta} \left[ \ln (1 + e^{\beta [\mu - vK_m]}) + \ln (1 + e^{\beta \mu}) \right] - \frac{v}{12\pi} K_m^3 f_0 (vK_m) \\ - \frac{vK_m^3}{6\pi} - \frac{\mu K_m^2}{4\pi} - \frac{\mu^3}{6\pi v^2} - \frac{\pi \mu (\kappa_B T')^2}{12v^2} \text{.}
	\end{split} \tag{3.18}
\end{equation}
Neglecting exponentially small terms, $\Omega$ is well approximated by
\begin{equation}
	\Omega \equiv \frac{-vK_m^3}{6\pi} - \frac{\mu K_m^2}{2\pi} - \frac{\mu^3}{6\pi v^2} - \frac{\pi \mu (\kappa_B T')^2}{12v^2} \text{.} \tag{3.19}
\end{equation}

To study the nondegenerate regime we employ Eq. (2.6) with the Fermi-Dirac distribution taken in its Maxwell-\newline Boltzmann limit,
\begin{equation}
	\frac{\partial f_0 (E)}{\partial E} = -\beta e^{\mu \beta} e^{-E \beta} \text{,} \tag{3.20}
\end{equation}
and then the grand potential $\Omega$ is given by
\begin{equation}
	\Omega = -\beta e^{\mu \beta} \int_0^\infty dE e^{-E\beta} \int_c \frac{ds}{2\pi i} e^{Es} \frac{\widehat{Z}(s)}{s^2} = -\beta^{-1} e^{\mu \beta} \widehat{Z}(\beta) \tag{3.21}
\end{equation}
(since the $E$- and $s$-integrals constitute a Laplace transform and its inverse). Employing Eq. (3.5), we have $\Omega$ in the nondegenerate regime as 
\begin{equation}
	\begin{split}
	\Omega = -\beta^{-1} e^{\mu\beta} \left\{ \frac{K_m^2}{4\pi} + \frac{K_m \sinh (v\beta K_m)}{\pi v\beta} - \frac{\cosh (v\beta K_m)}{\pi v^2 \beta^2} + \right. \\ \left. \frac{1}{\pi v^2 \beta^2} \right\} \text{.} 
	\end{split} \tag{3.22}
\end{equation}
It is of interest to note that the density $n$ is given in the nondegenerate regime by
\begin{equation}
	\begin{split}
	n = \int \frac{d\omega}{2\pi} \int \frac{d^2 K}{(2\pi)^2} f_0(\omega) A(\vec{K},\omega) = e^{\mu \beta} \int \frac{d\omega}{2\pi} \int \frac{d^2 K}{(2\pi)^2} e^{-\beta\omega} \\ \times A(\vec{K}, \omega) \text{,} \end{split} \tag{3.23}
\end{equation}
where $A(\vec{K},\omega)$ is the spectral weight embodied in the trace of the retarded Green's function as
\begin{equation}
	A(\vec{K},\omega) = -2 ImTr G^{\text{ret}} (\vec{K}, \omega) = 2\pi\delta (\omega) + 2\pi \sum_\pm \delta (\omega \pm vK) \tag{3.24}
\end{equation}
for the Dice lattice. Employing Eqns. (3.23) and (3.24), we have the density as
\begin{equation}
	n = e^{\mu\beta} \left\{ \frac{K_m^2}{4\pi} + \frac{1}{2\pi v^2 \beta^2} \sum_\pm \int_0^{\pm v\beta K_m} dE Ee^{-\beta E} \right\} \text{,} \tag{3.25}
\end{equation}
and evaluation of the integrals on the right yields
\begin{equation}
	n = e^{\mu\beta} \left\{ \frac{K_m^2}{4\pi} + \frac{K_m \sinh (vK_m \beta)}{\pi v\beta} - \frac{\cosh (vK_m \beta)}{\pi v^2 \beta^2} + \frac{1}{\pi v^2 \beta^2} \right\} \text{.} \tag{3.26}
\end{equation}
This provides the nondegenerate result for $\Omega$ in the form
\begin{equation}
	\Omega = -\beta^{-1} n = -n \kappa_B T' \text{;} \tag{3.27}
\end{equation}
which could have been anticipated because of the similar dependencies of the partition function $\widehat{Z}(\beta)$ and the density $n$ on the trace of the retarded Green's function.

\newpage
\noindent \textbf{4. Entropy and the Specific Heat}

The entropy may be determined using the thermodynamic relation$^{24}$
\begin{equation}
	dF = -PdV - SdT^{'} + \mu dN \tag{4.1}
\end{equation}
by variation holding both area (volume) and number fixed, with the result
\begin{equation}
	S = -\left(\frac{\partial F}{\partial T^{'}} \right)_{N,V,\mu} = -\left(\frac{\partial (F-\mu n)}{\partial T^{'}}\right)_{N,V,\mu} = -\left(\frac{\partial \Omega}{\partial T'}\right)_{N,V,\mu} \text{.} \tag{4.2}
\end{equation}
In the degenerate regime, Eq. (3.19) may be employed to obtain the Dice Lattice entropy, with the result
\begin{equation}
	S_{\text{Deg}} = \frac{\pi \mu k_B^2 T'}{6v^2} \tag{4.3}
\end{equation}
and the corresponding specific heat (at constant volume) in the degenerate regime is$^{25}$
\begin{equation}
	C_v = T' \left( \frac{\partial S}{\partial T'} \right)_V = \frac{\pi\mu k_B^2 T'}{6v^2} \text{.} \tag{4.4}
\end{equation}

To determine the entropy of the Dice lattice in the nondegenerate regime, we employ Eq. (3.27), $\Omega = -n\kappa_B T'$, and use Eqns. (3.24, 4.2) to obtain 
\begin{equation}
	\begin{split}
	S = -\left( \frac{\partial\Omega}{\partial T'} \right)_{N,v,\mu} = \kappa_B n = \kappa_B e^{\mu\beta} \left\{ \frac{K_m^2}{4\pi} + \frac{K_m \sinh (vK_m\beta)}{\pi v\beta}  \right. \\ \left. - \frac{\cosh (vK_m\beta)}{\pi v^2 \beta^2} + \frac{1}{\pi v^2 \beta^2} \right\} \text{.} 
	\end{split} \tag{4.5}
\end{equation} 
This yields the nondegenerate specific heat at constant volume as
\begin{equation}
	\begin{split}
	C_v = \kappa_B e^{\mu\beta} \left\{ \cosh (vK_m \beta) \left[ \frac{\mu}{\pi v^2 \beta} - \frac{2}{\pi v^2 \beta^2} - \frac{K_m^2}{\pi} \right] + \right. \\ \left. \sinh (vK_m \beta) \left[ \frac{2K_m}{\pi v\beta} - \frac{\mu K_m}{\pi v} \right] + \frac{2}{\pi v^2 \beta^2} - \frac{\mu}{\pi v^2 \beta} - \frac{\mu \beta K_m^2}{4\pi} \right\} \text{.} 
	\end{split} \tag{4.6}
\end{equation}
As noted in the introduction, knowledge of specific heat is pertinent to the characterization of materials and understanding their basic physical properties$^{12,13-17}$, including their ability to assist in the management of dissipated heat.

\newpage
\noindent \textbf{\LARGE{REFERENCES}} \newline \newline
1. J.D. Malcolm and E.J. Nicol, Phys. Rev. B 93, 165433 (2016). \newline
2. D. Bercioux, D. F. Urban, H. Grabert and W. H$\ddot{\text{a}}$usler, Phys. Rev. A 80, 063683 (2009). \newline
3. G. K. Ahluwalia, Editor: ``Applications of Chalcogenides: S, Se, Te," Springer (2017). \newline
4. S. Q. Shen, ``Topological Insulators," Springer (2012). \newline
5. M. J. S. Spencer, ``Silicene," Springer (2016). \newline
6. M. Katsnelson, ``Graphene: Carbon in Two Dimensions," Cambridge University Press (2012). \newline
7. H. Aoki and M. S. Dresselhaus, ``Physics of Graphene," Springer (2013). \newline
8. E. L. Wolf, ``Graphene: A New Paradigm in Condensed Matter and Device Physics,"
(2013). \newline
9. T. 0. Wehling, A. M. Black-Shaffer and A. V. Balatsky, ``Dirac Materials," arXiv: 1405.5774 vl [cond-mat.mtrl-sci] (22 May 2014). \newline
10. J. Wang, S. Deng, Zhongfan Liu and Zhirong Liu, ``The Rare 2D Materials with Dirac Cones," National Science Review 2:22-39 (2015). \newline
11. N. J. M. Horing, ``Aspects of the Theory of Graphene," Transactions Royal Society A 368, 5525-56 (2010). \newline
12. Stewart, G. R., ``Low Temperature Specific Heat of Layered Transition Metal Dichalcogenides," J. Superconductivity and Novel Magnetism 33, 213-215 (2019). \newline
13. DiSalvo, F.J., Schwall, R., Geballe, T.H., Gamble, F.R., Osiecki, J.H: Phys. Rev. Lett. \textbf{27}, 310 (1971). \newline
14. Meyer, S.F., Howard, R.E., Stewart, G.R., Acrivos, J.V., Geballe, T.H.: J. Chem. Phys. \textbf{62}, 4411 (1975). \newline
15. Shwall, R.E., Stewart, G.R., Geballe, T.H.: J. Low Temp. Phys. \textbf{22}, 557 (1976). \newline
16. Murphy, D.W., DiSalvo, F.J., Hull, G.W.J., Waszczak, J.V., Meyer, S.F., Stewart, G.R., Early, S., Acrivos, J.V., Geballe, T.H.: J. Phys. Chem. \textbf{62}, 967 (1975). \newline
17. Bachmann, R., DiSalvo, F.J., Geballe, T.H., Greene, R.L., Howard, R.E., King, C.N., Kirsch, H.C., Lee, K.N., Schwall, R.E., Thomas, H.U., Zubeck, R.B.: Rev. Sci. Instrum. \textbf{43}, 205 (1972). \newline 
18. A. H. Wilson, ``The Theory of Metals," 2$^\text{nd}$ Edition, Cambridge University Press, Section 6.6, (1965). \newline
19. A. Erdelyi, Editor, ``Tables of Integral Transforms I," McGraw-Hill p. 131, \#21 (1954). \newline
20. N. J. M. Horing, ``Quantum Statistical Field Theory," Oxford University Press (2017). \newline
21. A. Erdelyi, Editor, ``Tables of Integral Transforms I," McGraw-Hill, p. 30 \#2 (1954). \newline
22. M. L. Glasser, ``Scr.Mat.Fiz.," No. 498-541, p. 49-50 (1975). \newline
23. A. H. Wilson, ``The Theory of Metals," 2$^\text{nd}$ Edition, Cambridge University Press, Appendix A4 (1965). \newline
24. M. W. Zemansky, ``Heat and Thermodynamics," McGraw-Hill, (1951). \newline
25. L. D. Landau and E.M. Lifshitz, ``Statistical Physics, Part 1, 3$^{\text{rd}}$ Ed.," Pergamon Press, Eq. (13.5) (1980). \newline
26. P. C. Martin and J. Schwinger, Phys. Rev. 115, 1342 (1959).
		
\end{document}